\documentclass[lettersize,journal]{IEEEtran}

\usepackage{enumitem}
\usepackage{hyperref}
\usepackage{ulem} 
\AtBeginDocument{%
	\providecommand\BibTeX{{%
			Bib\TeX}}}
\usepackage{multirow}
\usepackage{colortbl} 
\usepackage{xcolor}
\usepackage{algorithmic}

\usepackage{amssymb}
\usepackage{booktabs}
\usepackage[linesnumbered,ruled,vlined]{algorithm2e}
\usepackage[most]{tcolorbox}
\usepackage{ulem}
\usepackage{threeparttable}
\def\BibTeX{{\rm B\kern-.05em{\sc i\kern-.025em b}\kern-.08em
    T\kern-.1667em\lower.7ex\hbox{E}\kern-.125emX}}
\usepackage{balance}

\begin{document}

\title{Boosting Vulnerability Detection with Inter-function Multilateral Association Insights\\}



\author{Shaojian Qiu, Mengyang Huang, Jiahao Cheng

\IEEEcompsocitemizethanks{


Shaojian Qiu, Mengyang Huang and Jiahao Cheng are with the College of Software Engineering, South China Agricultural University, Guangzhou, 510642, China (e-mails: qiushaojian@scau.edu.cn; huangmengyang@insoft-lab.com; chengjiahao@insoft-lab.com). 

}
}

\markboth{Journal of \LaTeX\ Class Files,~Vol.~18, No.~9, September~2020}%
{How to Use the IEEEtran \LaTeX \ Templates}
\maketitle

\begin{abstract}
Vulnerability detection is a crucial yet challenging technique for ensuring the security of software systems. Currently, most deep learning-based vulnerability detection methods focus on stand-alone functions, neglecting the complex inter-function interrelations, particularly the multilateral associations. This oversight can fail to detect vulnerabilities in these interrelations. To address this gap, we present an Inter-Function Multilateral Association analysis framework for Vulnerability Detection (IFMA-VD). The cornerstone of the IFMA-VD lies in constructing a code behavior hypergraph and utilizing hyperedge convolution to extract multilateral association features. Specifically, we first parse functions into a code property graph to generate intra-function features. Following this, we construct a code behavior hypergraph by segmenting the program dependency graph to isolate and encode behavioral features into hyperedges. Finally, we utilize a hypergraph network to capture the multilateral association knowledge for augmenting vulnerability detection. We evaluate IFMA-VD on three widely used vulnerability datasets and demonstrate improvements in F-measure and Recall compared to baseline methods. Additionally, we illustrate that multilateral association features can boost code feature representation and validate the effectiveness of IFMA-VD on real-world datasets.
\end{abstract}


\section{Introduction}
As the complexity of modern software systems continues to grow, the possibility of code vulnerabilities emerging also increases \cite{lin2020software,jiang2024stagedvulbert}. This trend is evident in data from the U.S. National Vulnerability Database, which report a 15\% year-over-year increase in vulnerabilities (totaling 28,828 in 2023), more than tripling the number reported a decade ago \cite{NVD2023}. 
Attackers can exploit these vulnerabilities to gain unauthorized access to systems, steal sensitive information, or disrupt critical operations. Therefore, effectively detecting potential vulnerabilities is a critical concern in the fields of software reliability and security \cite{ni2023function,nong2022open}.

Over the past few years, deep learning has gained widespread application in vulnerability detection tasks~\cite{nong2022open,li2021sysevr}. The typical deep learning-based detection approaches involve transforming code functions into various representations (e.g., text, abstract syntax tree (AST), code property graph (CPG)) and subsequently converting these representations into vectors for intra-function feature extraction using deep neural networks \cite{guo2022unixcoder, zhang2023vulnerability}. Then, these deep learning-generated features are employed to perform vulnerability detection \cite{wen2023meta, zhang2024evm}.

\begin{figure}[htbp]
    \centerline{\includegraphics[width=0.99\linewidth]{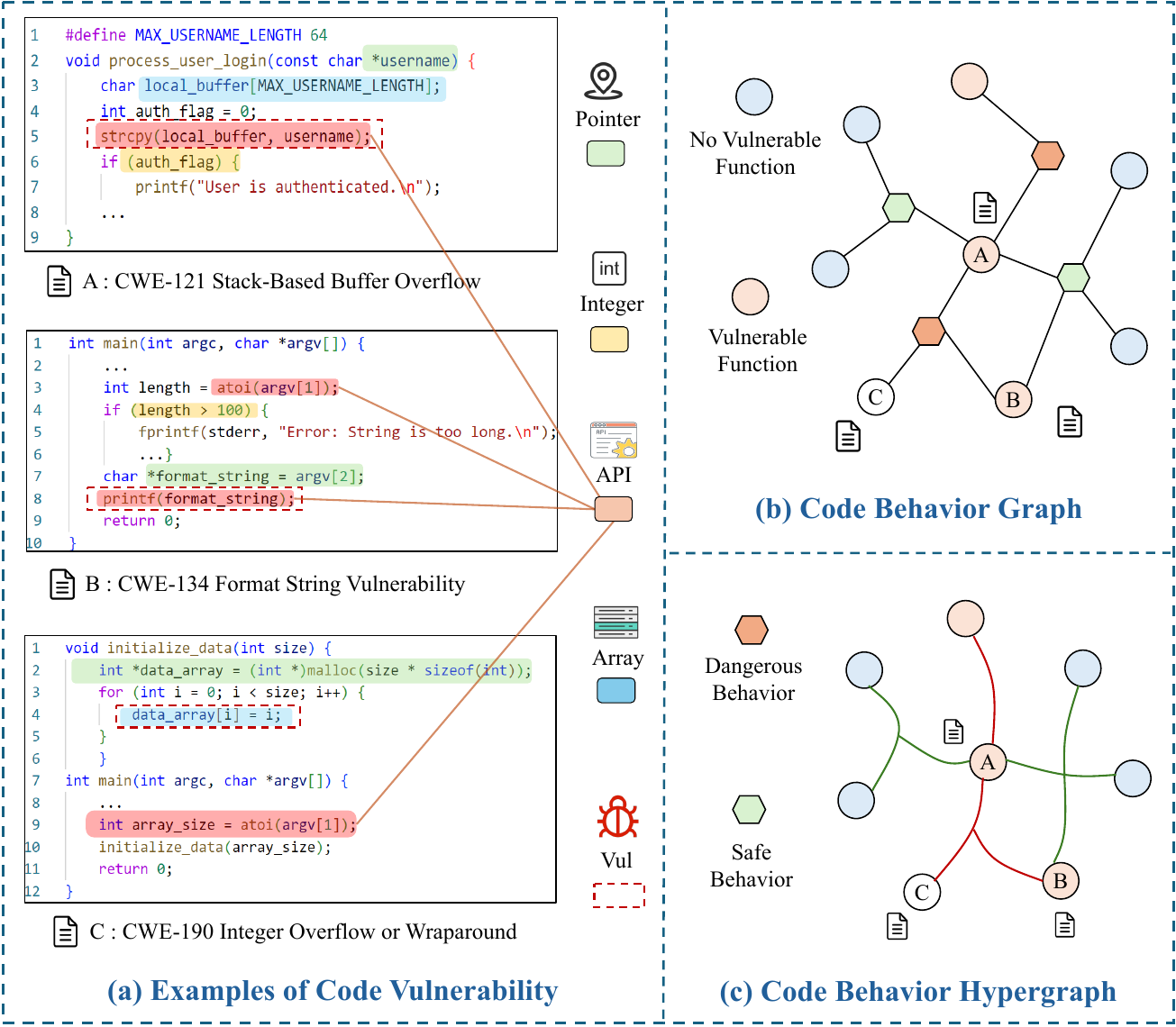}}
    \caption{Motivating Case.}
    \label{fig1}
\end{figure}

However, the above studies primarily focus on stand-alone functions, neglecting the relationships between them. Figure \ref{fig1} (a) illustrates an example of inter-function relationships, highlighting three common types of vulnerable functions: A, B, and C. The implementation of these functions relies on underlying algorithms, code elements, and programming patterns, collectively referred to as code behavior \cite{yuan2023enhancing}. In fact, different functions may exhibit similar code behaviors and can form relationships by sharing these behaviors. To apply these interrelations, some work \cite{yuan2023enhancing} has attempted to construct code behavior graphs to extract inter-function features for vulnerability detection. Although it shows promising performance, the related methods face two primary limitations: (1) high computational overhead due to the complexity of the code behavior graph and (2) inability to represent multilateral associations among functions.

To elucidate these limitations, we present a motivating case in Figure \ref{fig1}(b). Existing studies extract behaviors through program slicing and use them as nodes to establish binary associations between functions\cite{yuan2023enhancing,yadavally2024learning}. However, representing code behavior as nodes within graphs results in complex structures, thereby increasing the computational costs associated with graph analysis \cite{xia2021graph}. Additionally, when multiple functions share one or more code behaviors, they can form multilateral behavioral associations. Nevertheless, current methods using behavior nodes as intermediaries for information transfer cannot directly represent these inter-function multilateral associations. They necessitate multiple embedding layers to enable effective communication between adjacent functions, but excessive layering can lead to overly smoothed node features, thus reducing overall embedding performance \cite{zhang2022rethinking, yang2021graphformers}. As a result, these two limitations may hinder the detection of complex vulnerabilities that arise from multilateral interactions among functions.

To handle the above limitations, we propose a hypergraph modeling approach with degree-free hyperedges to effectively encapsulate the multilateral behavioral associations among functions \cite{di2022generating, feng2023hypergraph}. As illustrated in Figure \ref{fig1}(c), compared to traditional binary edges, we encode code behaviors within hyperedges, which significantly reduces the complexity of the graph and directly represents the multilateral behavioral associations among functions. 

Based on introducing the hypergraph concept, we innovatively propose an Inter-function Multilateral Association analysis framework for Vulnerability Detection (IFMA-VD). Specifically, this framework first captures a CPG by parsing functions and extracts intra-function features using a gated residual graph neural network. Next, IFMA-VD constructs a code behavior hypergraph by segmenting the program dependency graph with sensitive points to isolate and encode behavioral features into hyperedges. Finally, a hypergraph neural network is employed to extract multivariate relational features among functions for vulnerability detection. We evaluate IFMA-VD on three widely used vulnerability datasets to verify its effectiveness. Compared with seven state-of-the-art vulnerability detection methods, experimental results demonstrate that IFMA-VD achieves superior F-measure and Recall results.

In summary, the main contributions of this paper are as follows:

\begin{itemize}[itemsep=5pt,topsep=7pt] 
\item We pioneer using hypergraphs to model multilateral behavioral associations and design the IFMA-VD to enhance vulnerability detection performance.
\item We rigorously evaluate IFMA-VD on three widely used vulnerability datasets, demonstrating substantial improvements in F-measure and Recall compared to the baseline methods.
\item We propose a plug-and-play method for extracting inter-function features by constructing a code behavior hypergraph and using hyperedge convolution to capture multilateral associations. Empirical studies demonstrate that integrating our method improves the detection performance of various intra-function feature-based methods.
\end{itemize}
 
The remainder of this paper is structured as follows. Section II reviews related work. Section III presents the method of our IFMA-VD. Sections IV and V elaborate on the experiment design and results. Section VI discusses the threats to validity. Section VII concludes this paper.

\section{Related work}

\begin{figure*}[h]
    \centerline{\includegraphics[width=0.99\linewidth]{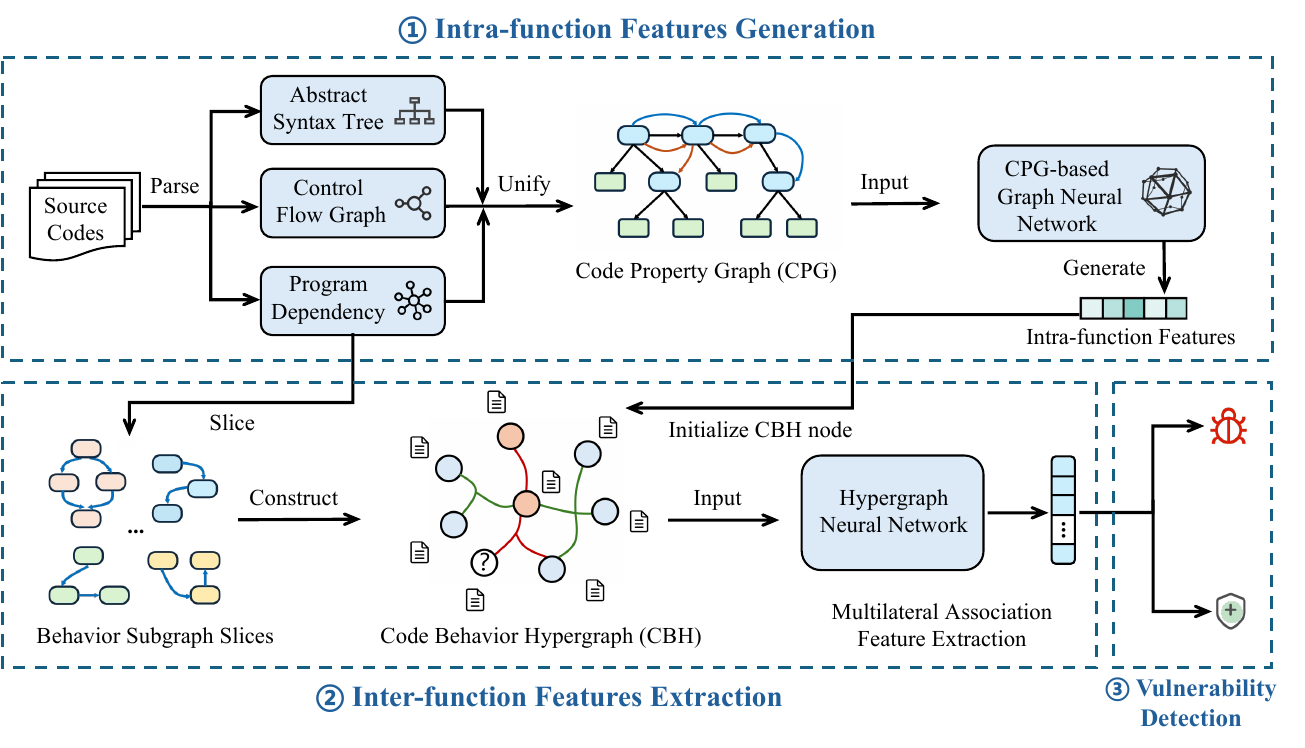}}
    \caption{The Framework of IFMA-VD}
    \label{fig2}
\end{figure*}

The vulnerabilities of software can result in unauthorized access, data breaches, or even a complete system compromise. Vulnerability detection is a crucial aspect of software security, allowing security teams to proactively identify potential weaknesses before attackers exploit them \cite{cheng2021deepwukong,mamede2022transformer}.   Traditional vulnerability detection methodologies are typically divided into two main categories: static analysis and dynamic analysis \cite{lipp2022empirical,wu2024ultravcs}. Static analysis examines software artifacts such as source code, bytecode, or binary code, enabling the early detection of vulnerabilities during development by leveraging predefined rules or patterns \cite{afrose2022evaluation,xue2020cross}. However, static analysis is limited by its reliance on human-crafted rules, which can result in high false positive rates and an inability to detect novel vulnerabilities \cite{liu2024semantic}. Dynamic analysis involves monitoring the program's execution in real time, using techniques like fuzz testing \cite{zhu2022fuzzing} and symbolic execution \cite{zheng2022park} to uncover vulnerabilities exposed during runtime. While dynamic analysis provides insights into runtime behavior, it may not explore all program execution paths and can be computationally expensive.

In recent years, deep learning has been applied to vulnerability detection to overcome the limitations of traditional detection methods and has demonstrated exceptional performance\cite{zhang2022example,chen2022ausera,zou2022mvulpreter}. These approaches can be broadly categorized into two main classes. (1) Text-based methods: Researchers have proposed treating code as natural language text and applying text classification methods based on NLP for vulnerability detection \cite{guo2022unixcoder,qiu2024vulnerability}. (2) Graph-based methods:  Unlike traditional text, code possesses rich semantic and structural information. Consequently, studies have transformed code into graph structures such as ASTs, program dependence graphs (PDGs), and CPGs through static analysis \cite{huang2024hevuld,cao2024coca,cao2022mvd}. These graph structures are then analyzed using GNN-based graph classification or node classification models to detect vulnerabilities. This approach leverages the expressive power of GNN to capture the complex relationships and dependencies inherent in software code, leading to more accurate vulnerability detection \cite{nong2022open,li2021vulnerability,hu2023interpreters}. Additionally, certain studies have delved into vulnerability detection by analyzing program data flow \cite{steenhoek2024dataflow} or employing Transformer-based pre-trained models \cite{liu2024pre}. Moreover, innovative research has converted code into image representations and applied computer vision techniques for vulnerability detection \cite{wu2022vulcnn}.

Existing methods primarily focus on stand-alone functions or  the information between statements, overlooking the underlying relationship among functions \cite{li2021sysevr, liu2024pre}. Understanding and leveraging the associations among functions is crucial for identifying vulnerabilities, as many security weaknesses arise from their multilateral interactions. Existing work by Yuan et al. \cite{yuan2023enhancing} proposes constructing code behavior graphs to explore binary associations between functions; however, this method exhibits high complexity and computational cost and cannot effectively express the multilateral associations. To address this challenge, our paper innovatively proposes leveraging hypergraphs to model the multilateral associations among functions. We have developed IFMA-VD, which enhances the performance of vulnerability detection.

\section{Method}

In this section, we introduce the IFMA-VD framework, which is centred on leveraging knowledge embedded in inter-function multilateral associations to enhance the performance of vulnerability detection. The architecture of the framework, depicted in Figure \ref{fig2}, comprises three primary phases: (1) parsing functions into CPGs and generating intra-function features, (2) constructing a code behavior hypergraph and extracting inter-function multilateral association features, and (3) applying classification techniques to detect the presence of vulnerabilities. The subsequent subsections will elaborate on the detailed implementation of each phase within the IFMA-VD framework.

\subsection{Intra-function Features Generation}

\begin{figure}[h]
    \centerline{\includegraphics[width=0.80\linewidth]{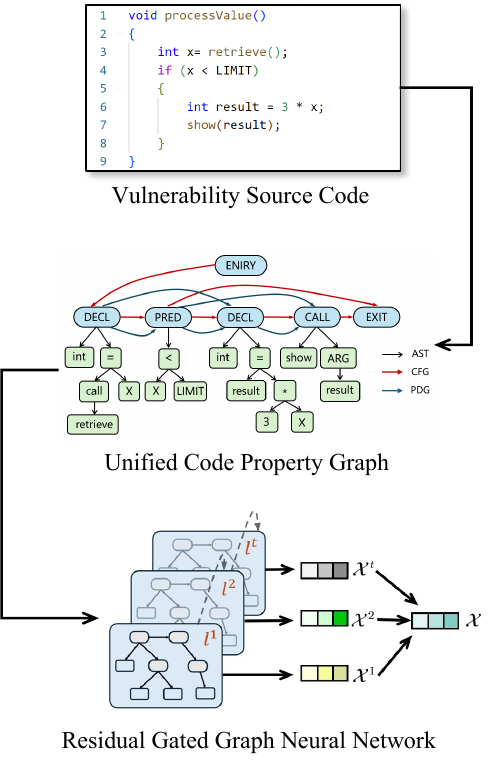}}
    \caption{The Process of Intra-function Features Generation}
    \label{fig3}
\end{figure}

In the initial phase of the IFMA-VD framework, code parsing is performed to generate intra-function features. These features encompass a variety of information (e.g., text, semantics, structure) within stand-alone functions, which facilitates a deeper representation of the code. As shown in Figure \ref{fig3}, we leverage the static code analysis tool, Joern, to parse functions and transform the code into a CPG representation. This representation effectively integrates ASTs, Control Flow Graphs (CFGs), and PDGs into a comprehensive code graph structure, which is widely adopted in software security and code representation tasks \cite{wen2023vulnerability,Devign}. Specifically, the AST captures the hierarchical relationship between syntactic elements, the CFG provides data flow information,  and the PDG indicates mutual dependencies among data. Joern parses these three graph types and integrates them into a unified representation, denoted as $\mathcal{G}_{cpg}(\mathcal{V}_{cpg}, \mathcal{E}_{cpg})$. Within this graph, nodes $\mathcal{V}_{cpg}$ represent code statements or syntactic elements, and edges $\mathcal{E}_{cpg}$ depict various types of information flow.

To generate intra-function features from $\mathcal{G}_{cpg}$, we implemented a feature generation module based on a gated graph neural network (GGNN), depicted in Figure \ref{fig3}. This module begins by embedding each node $n \in \mathcal{V}_{cpg}$ into a $d$-dimensional feature vector $u \in \mathbb{R}^d$ using Word2Vec, with all nodes collectively forming the feature matrix  $\mathcal{H}=\left[ u^1,u^2,\cdots ,u^n \right] ^T\in \mathbb{R} ^{n\times d}$. To enhance the contextual understanding of the node information within $\mathcal{G}_{cpg}$, an aggregation function $f_{GRU}$, based on the gated recurrent unit (GRU), is employed in the GGNN to capture the structural context of the code. The function operates as follows:
\begin{align}
\mathcal{H}^{t+1}=f_{GRU}\left( \mathcal{A}\mathcal{H}^t,\mathcal{H}^t \right) 
\end{align}
where $\mathcal{A}$ denotes the adjacency matrix of $\mathcal{G}_{cpg}$, and $\mathcal{H}^t$ represents the node feature matrix at the $t$-th layer. The updated node representation $\mathcal{H}^{t+1}$ is obtained by aggregating information, $\mathcal{A}\mathcal{H}^t$, from neighboring nodes through the GRU, with $\mathcal{H}^t$ serving as the input. Subsequently, an average pooling layer is employed to consolidate the features from all nodes across each layer, concatenating them to form the intra-function representation $x \in \mathbb{R}^d$. Features are generated for each function within the vulnerability dataset to compile the intra-function feature set $\mathcal{X} \left\{ x^1,x^2,...,x^N \right\}$. We divided the vulnerability dataset in an 8:1:1 ratio and trained the intra-function feature generation module. It is crucial to note that the entire IFMA-VD framework uses a consistent dataset split across all phases of training, thereby eliminating the risk of information leakage.

\subsection{Inter-function Features Extraction}

\begin{figure*}[h]
    \centerline{\includegraphics[width=0.99\linewidth]{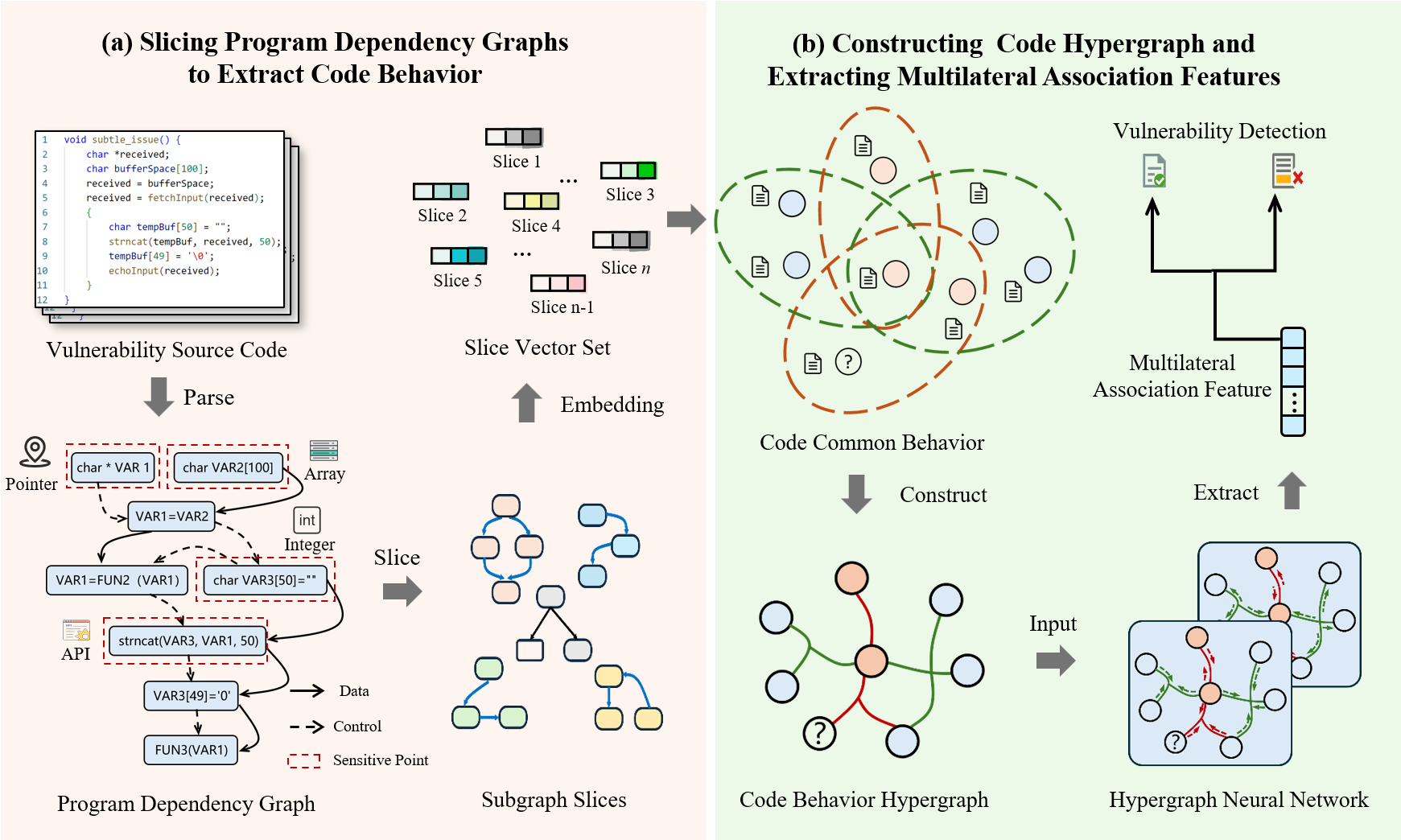}}
    \caption{The Process of Inter-function Features Extraction}
    \label{fig4}
\end{figure*}

\subsubsection{Code Behaviors Generation}

As illustrated in Figure \ref{fig4}, this phase of the IFMA-VD framework concentrates on constructing a code behavior hypergraph and extracting inter-function multilateral association features. Specifically, It encompasses two subphases: (1) generating code behavior through program slicing techniques, and (2) leveraging identified common behaviors among functions to construct a code hypergraph and extract multilateral association features. This method capitalizes on the underlying similarities and interactions among functions to identify potential security vulnerabilities more effectively.

In software development, a function is conceptualized as a set of code designed to perform specific tasks. Although different functions fulfill distinct purposes, they frequently exhibit common logic, algorithms, and programming patterns in their subtasks, collectively called behavior. The cornerstone of the IFMA-VD framework lies in constructing multilateral associations among functions based on these shared code behaviors, thereby enhancing vulnerability detection.

Initially, as shown in Figure \ref{fig4} (a), the source code is parsed into a PDG using Joern, denoted as $\mathcal{G}_{pdg}\left(\mathcal{V}_{pdg},\mathcal{E} _{pdg}\right)$. In this graph, the set of nodes  $\mathcal{V} _{pdg}$  corresponds to code statements, whereas the set of edges $\mathcal{E} _{pdg}$ represents control dependencies and data dependencies. This introduces the concept of vulnerability interest points, which are specific locations in the code identified through expert analysis as prone to vulnerabilities. Research has shown that sensitive APIs, as well as locations involving arrays, integers, and pointers, are common sources of vulnerabilities in code.  We then identify points of vulnerability interest within $\mathcal{G}_{pdg}$ as the starting points for slicing. Specifically, the process first identifies whether the nodes in the $\mathcal{G}_{pdg}$, representing code statements, correspond to vulnerability interest points. Then, forward and backward slicing is performed on these interest points, and the slicing results are concatenated to form behavior sub-graphs $\mathcal{G} _{pdg}^{sub}$. Each slice generated in this step contains information about the data flow and control flow within the context of these sensitive points, including all possible execution paths from the program entry point. This method effectively isolates statements unrelated to vulnerabilities while preserving the structural information of the code context. For all functions, a set of subgraphs $\mathcal{F}_{subg}$ is generated, each corresponding to different behaviors, as follows:
\begin{align}
\mathcal{F}_{subg}\left\{ \mathcal{G} _{pdg}^{sub^1}, \mathcal{G} _{pdg}^{sub^2},\cdots ,\mathcal{G} _{pdg}^{sub^N} \right\} 
\end{align}

Finally, all subgraphs within $\mathcal{F}_{subg}$ are initialized using Word2Vec and subsequently embedded into the vector set $\mathcal{F}_b = \left\{ b^1, b^2, \dots, b^N\right\}$, where each $b \in \mathbb{R}^d$ represents a code behavior vector. This embedding process is facilitated by the GGNN-based feature generation module trained in the previous phase, ensuring the vectors effectively encapsulate the structural and contextual nuances of the code behaviors. By embedding these subgraphs into a unified vector space, we create a foundation for analyzing the multilateral associations among different code behaviors, which is crucial for subsequent steps.

\subsubsection{Constructing Code Hypergraph and Extracting Multilateral Association Features}
As shown in Figure \ref{fig4} (b), this phase involves constructing a hypergraph of behaviors among functions to extract multilateral association features. Similar code behaviors often exist between different functions, forming complex multilateral association networks. By analyzing these associations, vulnerabilities with a common root cause can be more efficiently identified. First, leveraging clustering techniques, we partition the set $\mathcal{F}_b$ into distinct clusters of common behaviors, denoted as $\mathcal{C}\left\{c^1, c^2, \ldots, c^n\right\}$.  Each cluster $c\left( b^1, b^2,\cdots ,b^n \right) $ encapsulates common behaviors, which aids in establishing multilateral associations among functions through shared behavioral patterns. To more effectively represent the associations among functions, we introduce a hypergraph structure for modeling. Unlike traditional binary graphs, where edges connect only pairs of nodes, hyperedges in a hypergraph can connect multiple nodes simultaneously, making them especially suitable for representing and analyzing complex, nonlinear multilateral data dependencies \cite{qiu2024code,feng2023hypergraph}. In our framework, we conceptualize a hypergraph as $ \mathcal{G}_{Hg} = (\mathcal{V}_{Hg}, \mathcal{E}_{Hg}, \mathcal{W})$, where $\mathcal{V}_Hg$ denotes the vertex set, $\mathcal{E}_{Hg}$ represents the hyperedge set, and $\mathcal{W}$ is a diagonal matrix that signifies the hyperedge weights.  The hypergraph's structure is characterized by a hyperedge incidence matrix $\mathcal{H}_{hg}$, with dimensions $|\mathcal{V}_{Hg}| \times |\mathcal{E}_{Hg}|$. Each element  $\hbar $ in $\mathcal{H}_{hg}$ is defined as:
\begin{align}
    \hbar \left( \nu,\varepsilon \right) =\begin{cases}1, if\,\,\nu\in \varepsilon\\0, if\,\,\nu \notin \varepsilon \\\end{cases}
\end{align}
where $\nu$ and $\varepsilon$ denote vertex and hyperedge, respectively. We can derive the vertex degree matrix   $\mathcal{D}_\nu $ and the hyperedge degree matrix $\mathcal{D}_\varepsilon$ from $\mathcal{H}_{hg}$. These represent the diagonal degree matrices of hyperedges and nodes, obtained by:
\begin{align}
\left\{ \begin{array}{l}
	\mathcal{D} _{\varepsilon}=\left[ \delta \left( \varepsilon _i \right) \right] ,i\in [1,|\mathcal{E} _{Hg}|]\\
	\delta \left( \varepsilon _i \right) =\sum_{\nu _j\in \mathcal{V}}{h}\left( \nu _j,\varepsilon _i \right)\\
	\mathcal{D} _{\nu}=\left[ d\left( \nu _j \right) \right] ,j\in [1,|\mathcal{V} _{Hg}|]\\
	d\left( \varepsilon _j \right) =\sum_{v_j\in e_i,e_i\in \mathcal{E}}{w}\left( \varepsilon _i \right) h\left( v_j,\varepsilon _i \right)\\
\end{array} \right. 
\end{align}

 In our methodology, each node in the hypergraph is considered a function, and we utilize the intra-function feature $\mathcal{X}$,  generated in the initial phase, to initialize the vertex set $\mathcal{V}_{Hg}$. Subsequently, the common code behaviors set is encoded into the hyperedges set $\mathcal{E}_{Hg}$. 
Each common behavior $c\left( b^1, b^2,\cdots ,b^n \right) $ is treated as a hyperedge, connecting functions that share behavior vectors within $c$. Ultimately, we construct a behavior hypergraph $\mathcal{G}_{Hg} = (\mathcal{V}_{Hg}, \mathcal{E}_{Hg}, \mathcal{W})$ that encapsulates the multilateral associations among functions. To extract multilateral association features, we employ a hypergraph neural network that uses hyperedge convolutional layers to capture higher-order information.

Initially, after deriving $\mathcal{H}_{hg}$, which encapsulates the multilateral association, the hypergraph Laplacian $\vartheta$, representing the normalized positive semi-definite Laplacian matrix of the constructed hypergraph, can be computed as follows:
\begin{align}
\vartheta  ={\mathcal{D}_\nu }^{-1/2}\mathcal{H}_{hg}\mathcal{WD}_{\varepsilon}^{-1}H^T\mathcal{D}_\nu^{-1/2} 
\end{align}
we define  $ \varPsi   =\mathcal{I}- \vartheta$, where $\mathcal{I}$ denotes the identity matrix.  Given a semi-positive definite matrix $ \varPsi  $,  we employ eigen decomposition  $ \varPsi   =\phi   \varOmega \phi   $ to derive orthonormal eigenvectors  $\phi   =\mathrm{diag}\left( \phi _1,\cdots,\phi _n \right) $ and the non-negative eigenvalue diagonal matrix  $\varOmega =\mathrm{diag}\left( \gamma _1,\cdots,\gamma _n \right) $.

The instance features $\mathcal{X}$ in the hypergraph is given by $\mathcal{\hat{X}}=\phi   ^T\mathcal{X}$, where eigenvectors acting as the Fourier bases and eigenvalues represent the frequencies. The spectral convolution of the feature $\mathcal{X}$ and filter $f$  can be formulated as follows:
\begin{align}
    f\ast \mathcal{X}=\phi   \left( \left( \phi   ^Tf \right)  \otimes   \left( \phi   ^T \mathcal{X} \right) \right) = \phi   f\left( \varOmega \right) \phi   ^T \mathcal{X}
\end{align}
where $ \otimes  $ represents  the element-wise product and  $f\left( \varOmega \right) =\mathrm{diag}\left( f\left( \gamma _1 \right),\cdots,f\left( \gamma _n \right) \right) $  is defined as a function of the Fourier coefficients. To reduce computational complexity and prevent overfitting, for $n$ instance features of dimension $\mathcal{C}_1$, with $\mathcal{X} \in \mathbb{R}^{n \times \mathcal{C}_1}$ in the hypergraph, the hyperedge convolution can be expressed as follows:




\begin{align}
\mathcal{Z}=ReLu({\mathcal{D}_\nu}^{-1/2}\mathcal{H}_{hg}\mathcal{WD}_{\varepsilon}^{-1}\mathcal{H}_{hg}^T\mathcal{D}_\nu^{-1/2}\mathcal{X}\beta  )	\label{Z}
\end{align}

The filter $\beta$ is applied uniformly across the hypergraph nodes. After the hyperedge convolution, we obtain the inter-function multilateral information $\mathcal{Z}\in \mathbb{R} ^{n\times C2}$. The hypergraph convolution layer facilitates transformative information exchange between node-edge nodes, using the hypergraph structure to effectively extract multilateral associations among functions. In our approach, the weight matrix $\mathcal{W}$ is designated as the identity matrix $\mathcal{I}$, suggesting the weights of all hyperedges are equal, and $\mathcal{C}1 = \mathcal{C}2$. Multiple hyperedge convolutional layers, incorporating the $ReLu$ activation function, constitute the proposed hypergraph convolutional neural network. This architecture effectively captures multilateral associations between functions, thereby enhancing code feature representation.

\subsection{Vulnerability Detection}
The primary objective of the IFMA-VD framework is to determine whether a function is vulnerable, constituting a standard binary classification task. The ultimate output, denoted as $\mathcal{Z}$, captures the multilateral association features of each function, derived from the hypergraph convolution network. These features represent complex interactions and dependencies among behaviors, crucial for identifying potential vulnerabilities. The final representation $\mathcal{Z}$ is processed through logistic regression to generate the probability of an input function having a vulnerability. Throughout the model training phase, cross-entropy loss is utilized to guide the optimization of IFMA-VD by measuring the difference between the model's predicted probabilities and the actual labels. The target is to minimize this cross-entropy loss, ensuring the model's predictions are as accurate as possible. During the testing phase, each prediction involves adding a small batch or a single function to be tested, which does not significantly affect the original code behavior hypergraph structure. Similar to the training process, all functions to be detected are processed and integrated into the code behavior hypergraph to extract multilateral association features. This batch processing approach enables efficient handling of a large number of detections.

\begin{algorithm}[t]
    \caption{IFMA-VD}
    
    \label{alg1:IFMA-VD}
      \small
    \renewcommand{\algorithmicrequire}{\textbf{Input:}}
    \renewcommand{\algorithmicensure}{\textbf{Output:}}
    \begin{algorithmic}[1]
    \setlength{\baselineskip}{1.2\baselineskip}
        \REQUIRE \quad\\
        Function set $\mathcal{S}$. \\
        \ENSURE \quad\\
        Vulnerability detection results. \\
        \hspace*{-0.25in} {\bf Step 1: Intra-function Code Feature Generation}\\
        \FOR{each function $s$ in $\mathcal{S}$}
            \STATE Parse function $s$ to obtain $\mathcal{G}_{ast}$, $\mathcal{G}_{cfg}$, and $\mathcal{G}_{pdg}$.
            \STATE Unify $\mathcal{G}_{ast}$, $\mathcal{G}_{cfg}$, and $\mathcal{G}_{pdg}$ into $\mathcal{G}_{cpg}$.
        \ENDFOR
        \STATE Train intra-function feature generation model.
        \STATE Generate feature set $\mathcal{X}$ from $\mathcal{G}_{cpg}$.
        
        \hspace*{-0.25in} {\bf Step 2: Inter-function Features Extraction}\\
        \STATE Initialize code behavior hypergraph $\mathcal{G}_{hg} = \left(\mathcal{V}_{hg}, \mathcal{E}_{hg}, \mathcal{W}\right)$.
        \FOR{each function $s$ in $\mathcal{S}$}
            \STATE Slice $\mathcal{G}_{pdg}$ to obtain behavior subgraphs $\mathcal{G}_{pdg}^{sub}$.
            \STATE Embed $\mathcal{G}_{pdg}^{sub}$ into code behaviors set $\mathcal{F}_b$.
        \ENDFOR
        \STATE Filter $\mathcal{F}_b$ to generate a common behavior set $\mathcal{C}$.
        \STATE Encode $\mathcal{C}$ into hyperedges $\mathcal{E}_{hg}$ and initialize nodes $\mathcal{V}_{hg}$ with features set  $\mathcal{X}$.
        
        \STATE Construct hypergraph neural network (HGNN).
        \STATE Train HGNN and extract inter-function features $\mathcal{Z}$, as \\described in Eq.\ref{Z}.
        
        \hspace*{-0.25in} {\bf Step 3: Vulnerability Detection}\\
        \STATE Classify $\mathcal{Z}$ to detect the presence of vulnerabilities.
    \end{algorithmic}
\end{algorithm}

Algorithm \ref{alg1:IFMA-VD} presents a pseudocode representation to improve the clarity and detail of the IFMA-VD process. In Step 1, we parse the functions to generate intra-function code features, which are the foundation for subsequent steps. In Step 2, IFMA-VD initially slices each function's PDG to extract code behaviors and construct a code hypergraph structure. Subsequently,  a hypergraph neural network mines multilateral association features among the functions. In the final step,  classification techniques are applied to detect the presence of vulnerabilities accurately.

\section{Experiment Design}
To verify the effectiveness of IFMA-VD, we focus on the following three research questions (RQs):

\textbf{RQ1}: How does the effectiveness of IFMA-VD in vulnerability detection compare to baseline methods?

\textbf{RQ2}: Do the proposed multilateral association features enhance the performance of vulnerability detection?

\textbf{RQ3}: Does IFMA-VD effectively detect the real world vulnerabilities?


In this section, we present the experiment datasets, baseline methods and evaluation metrics.

\subsection{Experiment Datasets}
To evaluate the effectiveness of IFMA-VD, we implemented experiments using three datasets extensively recognized in vulnerability detection research: FFmpeg, Qemu, and Chrome+Debain. The statistics of the datasets are illustrated in Table \ref{table1}. FFmpeg and Qemu are balanced vulnerability datasets that consist of open-source software written in C. Specifically, FFmpeg, a multimedia framework, exhibits a vulnerability rate of 51.1\%, whereas Qemu, a machine emulator and virtualizer, has a vulnerability rate of 42.6\%. The Chrome+Debian dataset, provided by ReVeal \cite{chakraborty2021deep}, is an unbalanced vulnerability dataset derived from two open-source projects: the Linux Debian Kernel and Chromium. It comprises 2,240 vulnerable entries and 20,494 non-vulnerable entries, with a vulnerability rate of 9.9\%. By conducting experiments across these varied datasets, each with distinct characteristics and vulnerability rates, we aimed to comprehensively assess the adaptability of IFMA-VD across different software environments. Each dataset was randomly split, allocating 80\% for the training set, 10\% for the validation set, and 10\% for the testing set.

\begin{table}[h]
  \centering
  \caption{Statistics of the Datasets.}
 \renewcommand{\arraystretch}{1.2}
    \begin{tabular}{c|c|c|c|c}
    \toprule
    \textbf{Datasets} & \textbf{Samples} & \textbf{Vul}   & \textbf{Non-vul} & \textbf{Vul Ratio} \\
    \midrule
    FFmpeg & 9,768  & 4,981  & 4,788  & 51.10\% \\
    Qemu  & 17,549 & 7,479 & 10,070 & 42.62\% \\
    Chrome+Debian & 22,734 & 2,240 & 20,494 & 9.85\% \\
    \bottomrule
    \end{tabular}%
  \label{table1}%
\end{table}%

\subsection{Baseline Methods}

In our evaluation, we compare IFMA-VD with seven state-of-the-art vulnerability detection methods. The comparison methods and experiment implementation details are as follows:

	\textbf{Devign:} Devign is a graph-based vulnerability detection method that constructs a joint graph integrating AST, CFG, DFG, and code sequences, utilizing graph neural network to detection \cite{Devign}.
 
       \textbf{CodeBERT:} In recent years, BERT-based technology has gained widespread use in code representation learning. As a comparison method, we chose the representative approach CodeBERT, a bimodal pre-trained model designed for both programming language and natural language \cite{feng2020CodeBERT}.

       \textbf{IVDect:} A method that identifies vulnerabilities by analyzing vulnerable statements and their surrounding context through code's data and control dependencies \cite{li2021vulnerability}.
       
  \textbf{Reveal:} Reveal is a vulnerability detection that utilizes a CPG and GGNN for embedding and incorporates the SMOTE algorithm to address issues of sample imbalance \cite{chakraborty2021deep}.
  
    \textbf{VulCNN:} A detection method based on code visualization that converts PDG into image representations and utilizes convolutional networks to extract code features \cite{wu2022vulcnn}.

    \textbf{PDBert:} A model that enhances vulnerability detection by leveraging pre-training on control and data dependency prediction tasks to improve code understanding \cite{liu2024pre}.
    
    \textbf{VulBG:} A method that enhances vulnerability detection performance by extracting code behaviors through program slicing to construct binary behavior graphs \cite{yuan2023enhancing}.

These seven baseline methods encompass a variety of approaches: token-based (CodeBERT and PDBert), graph-based (Devign and Reveal), code visualization-based (VulCNN), and code behavior graph-based (VulBG). By comparing IFMA-VD with these methods, we can comprehensively evaluate the effectiveness of the IFMA-VD framework in vulnerability detection.

\textbf{Implementation Details of IFMA-VD:} 
To ensure the fairness of our experiments, we applied the same data split across all compared methods, randomly splitting the dataset into training, validation, and testing sets in an 8:1:1 ratio. We implemented some of the compared methods in the experiments (e.g., VulCNN \cite{wu2022vulcnn} and Devign \cite{Devign}) using the source code available online. For baseline methods that do not provide source code, we ensured implementation by strictly following the details provided in the original papers. The hyperparameters of the IFMA-VD model proposed in this paper encompass the dimensions of intra-function features and multilateral association features derived from hyperedge convolution, the number of hyperedges $K$, and the network training parameters. Specifically, $K$ is configured to 1,000, and the feature dimension is set to 256. For the learning parameters, the epoch count is set to 200, and the learning rate is set to 0.01. The Adam optimizer is employed, with gamma set to 0.9 and weight decay to 5e-4.

\subsection{Evaluation Metrics}
To evaluate detection performance, we employ the widely used metrics Recall, and F-measure in vulnerability detection research. In vulnerability detection tasks, there are four possible outcomes: detecting an instance with a true vulnerability as vulnerable (True Positive: TP), detecting a clean instance as vulnerable (False Positive: FP), detecting an instance with a true vulnerability as clean (False Negative: FN), and detecting a clean instance as clean (True Negative: TN). The evaluation metrics Recall, and F-measure are calculated as follows: 
\begin{align}
\text{Recall} &= \frac{\text{TP}}{\text{TP} + \text{FN}} 
\end{align}
\begin{align}
\text{F-measure} &= \frac{2 \times \text{TP}}{2 \times \text{TP} + \text{FP} + \text{FN}}
\end{align}

Recall quantifies the ratio of TP instances to the total number of actual vulnerable instances. A higher Recall value indicates that the model has a stronger ability to detect vulnerabilities. The F-measure, which is the harmonic mean of Precision and Recall, offers a comprehensive assessment of the model's performance in vulnerability detection.

\section{Experiment results}

\subsection{\textbf{RQ1: How Does the Effectiveness of IFMA-VD in Vulnerability Detection Compare to Baseline Methods?}}
To answer the first question, we evaluated the effectiveness of IFMA-VD in vulnerability detection by comparing its performance to seven baseline methods across three datasets. The results are shown in Table \ref{baselines}, with the best performance in bold and the second-best underlined with a double line. Specifically, IFMA-VD achieved F-measures of 69.1, 62.5, and 50.8 across the three datasets. These results represent improvements of 5.5\%-33.1\%, 1.0\%-26.8\%, and 6.1\%-81.1\% compared to the other methods. This substantial improvement in the F-measure indicates that IFMA-VD effectively balances Precision and Recall, making it a reliable method for vulnerability detection.


In terms of Recall, which measures the ability to identify true positive instances, IFMA-VD obtained values of 95.9, 74.4, and 66.8 across the three datasets. These recall values reflect enhancements of 16.4\%-87.3\%, 15.2\%-42.3\%, and 32.5\%-186.9\% over the baseline methods. The high recall rates indicate that IFMA-VD can detect vulnerabilities, ensuring fewer vulnerabilities are missed during the detection process.

The final row of Table \ref{baselines} summarizes the Win/Tie/Loss (W/T/L) results, comparing IFMA-VD's performance against each baseline method on individual metrics within each dataset. For example, a W/T/L result 7/0/0 for the F-measure on the FFmpeg dataset indicates that IFMA-VD outperformed all seven baseline methods with no ties or losses. Overall, IFMA-VD demonstrated superior performance in most vulnerability detection tasks, achieving W/T/L results of 7/0/0 in F-measure and Recall metrics. These consistent wins across various metrics and datasets underscore IFMA-VD's effectiveness.

\begin{table*}[htbp]
  \centering
  \caption{Performance of IFMA-VD v.s. Baseline Methods (for RQ1).}
   \renewcommand{\arraystretch}{1.2}

    \begin{threeparttable}
    \begin{tabular}{c|cc|cc|cc}
    
    \toprule
    \multirow{2}[4]{*}{\textbf{Method}} & \multicolumn{2}{c|}{\textbf{FFmpeg}} & \multicolumn{2}{c|}{\textbf{Qemu}} & \multicolumn{2}{c}{\textbf{Chrome+Debian}} \\
\cmidrule{2-7}          & \textbf{F-measure} & \textbf{Recall} & \textbf{F-measure} & \textbf{Recall} & \textbf{F-measure} & \textbf{Recall} \\
    \midrule
    Devign & 51.9  & 52.1  & 53.7  & 53.4  & 28.4  & 28.7 \\
    CodeBERT & 53.0  & 51.2  & 54.1  & 52.3  & 25.4  & 27.4  \\
    IVDect & \uuline{65.5}  & 64.6  & 57.9  & \uuline{64.6}  & 38.8  & 39.5 \\
    Reveal & 62.6  & \uuline{82.4}  & 49.3  & 54.0  & 26.3  & 28.6  \\
    VulCNN & 54.2  & 57.7  & 55.1  & 58.2  & 31.5  & 51.0  \\
    PDBert & 52.4  & 43.2  & \uuline{62.4}  & 59.8  & \uuline{47.9}  & 45.4 \\
    VulBG & 57.5  & 62.1  & 55.9  & 58.9  & 36.5  & \uuline{59.3}  \\
    \midrule
    IFMA-VD(ours) & \textbf{69.1} & \textbf{95.9} & \textbf{62.5} & \textbf{74.4} & \textbf{50.8} & \textbf{78.6} \\
    \midrule
    W/T/L & 7/0/0 & 7/0/0 & 7/0/0 & 7/0/0 & 7/0/0 & 7/0/0 \\
    \bottomrule
    \end{tabular}%
  
    \end{threeparttable}
    
  \label{baselines}%
\end{table*}%

~\\
		\noindent
		\begin{minipage}[b]{0.45\textwidth}
			\fbox{
				\parbox{\textwidth}{
					\textbf{Answer to RQ1:} 
     The experimental results demonstrate the effectiveness of IFMA-VD in vulnerability detection tasks. Compared to seven state-of-the-art methods, IFMA-VD shows significant improvements in F-measure and Recall. These results confirm that IFMA-VD offers substantial advantages in accurately detecting vulnerabilities while maintaining balanced performance across evaluation metrics.
				}
			}
		\end{minipage}
  ~\\


\subsection{\textbf{RQ2: Do the Proposed Multilateral Association Features Enhance the Performance of Vulnerability Detection?}}

To address this research question, we investigated the contribution of multilateral association features to the performance of IFMA-VD and other vulnerability detection methods. This section compares three groups of vulnerability detection methods:

\textbf{Vulnerability Detection with Intra-function Features}: Existing vulnerability detection methods (e.g., Devign, VulCNN, Reveal) primarily focus on intra-function features, neglecting associative information among functions. These intra-function feature extraction methods form the first comparison group.

\textbf{Vulnerability Detection with Inter-function Features of Binary Association}: Recent studies have proposed constructing code behavior graphs (BGs) to extract inter-function features. However, this approach is highly complex and fails to capture multilateral associations among functions. These methods constitute the second comparison group.

\textbf{Vulnerability Detection with Inter-function Features of Multilateral Association}: To address the limitations of existing studies, we propose constructing code behavior hypergraphs (BHG) through program slicing and utilizing hypergraph neural networks to extract multilateral association features among functions, thereby boosting vulnerability detection. This approach forms the third comparison group.

Specifically, we selected four existing methods, Devign, CodeBERT, Reveal, and VulCNN, along with the first phase of IFMA-VD proposed in this paper, making a total of five intra-function feature extraction methods as the baseline methods in the first group, denoted as "base."  We then extended these baseline methods by constructing BGs to extract inter-function features, forming the second group, denoted as "+BG" (e.g., Devign+BG). Finally, we further extended these baseline methods by constructing code behavior hypergraphs and using hypergraph neural networks to extract multilateral association features, forming the third group, denoted as "+BHG" (e.g., Devign+BHG).

\begin{figure}[h]
    \centering

    \vspace{5pt}
    
    \begin{minipage}[t]{0.40\textwidth}
        \centering
        \includegraphics[width=\textwidth]{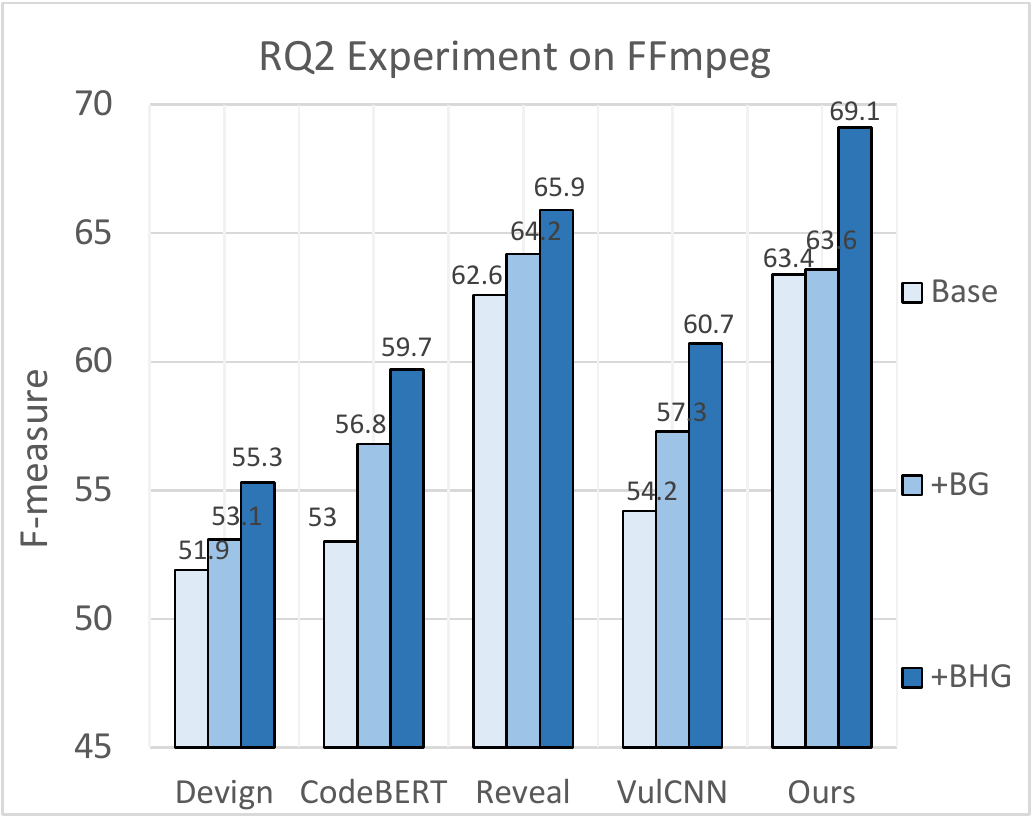}
    \end{minipage}
    
    \vspace{5pt}
    
    \begin{minipage}[t]{0.40\textwidth}
        \centering
        \includegraphics[width=\textwidth]{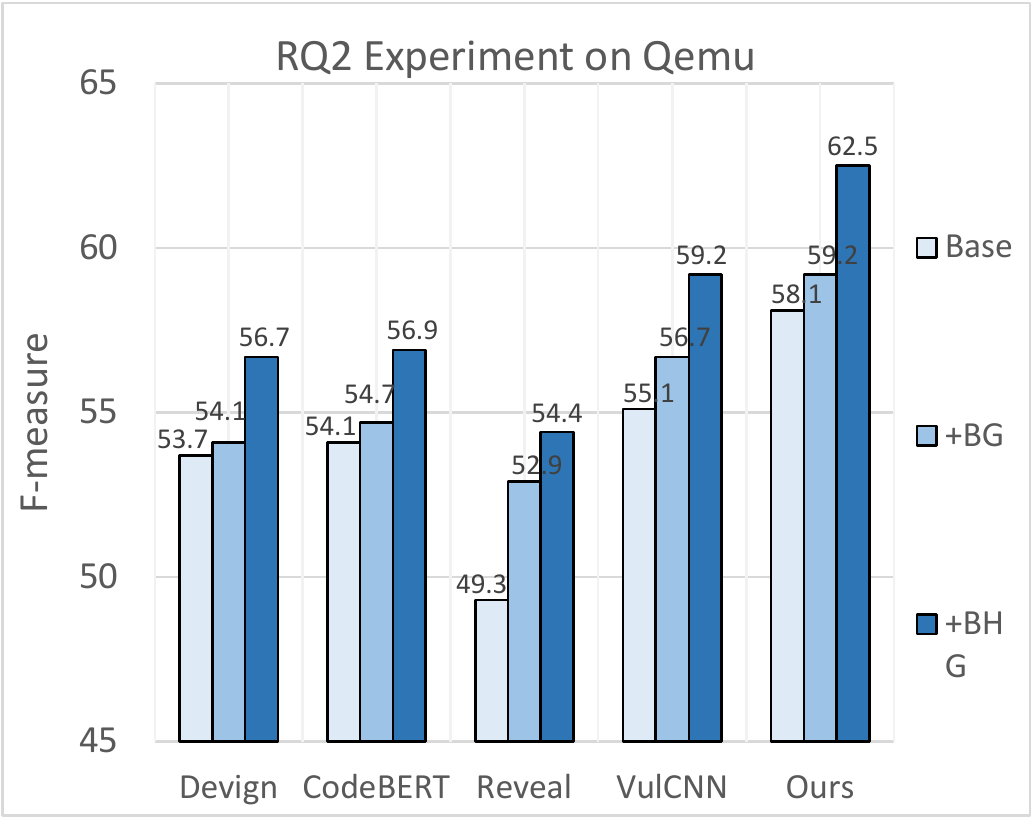}
    \end{minipage}
    
    \vspace{5pt}
    
    \begin{minipage}[t]{0.40\textwidth}
        \centering
        \includegraphics[width=\textwidth]{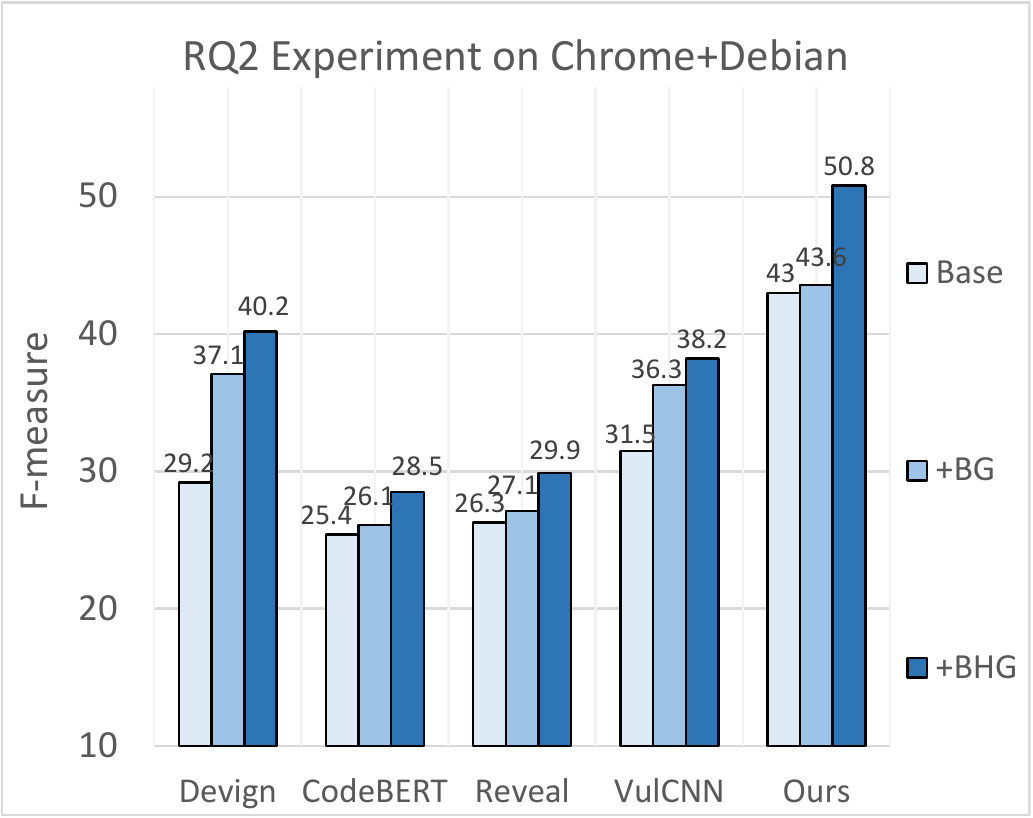}
    \end{minipage}
    \caption{F-measure Bar Chart of Three Comparison Groups on FFmpeg, Qemu, and Chrome+Debian Datasets (for RQ2).}
   \label{RQ2}
\end{figure}

Figure \ref{RQ2} presents a bar chart visualizing the F-measure achieved on three datasets by the baseline methods ("base") and their "+BG" and "+BHG" extensions. As shown in Figure \ref{RQ2}, extracting inter-function features by constructing BG enhances detection performance on the five base methods, with improvements ranging from 0.3\% to 7.2\% on FFmpeg, 0.7\% to 7.3\% on Qemu, and 1.4\% to 27.1\% on Chrome+Debian. To address the limitations of "base" and "+BG" methods, this paper proposes constructing "+BHG" to leverage hypergraph neural networks for mining multilateral association features, further enhancing vulnerability detection. The results show that our approach significantly improves the five base methods, with increases ranging from 5.3\% to 12.6\% on FFmpeg, 5.2\% to 10.3\% on Qemu, and 7.0\% to 37.7\% on Chrome+Debian. Additionally, our proposed IFMA-VD shows considerable improvements compared to the "+BG" methods.

Overall, the experimental results underscore the benefits of incorporating multilateral association features into the IFMA-VD framework. Additionally, these results indicate that IFMA-VD is a plug-and-play method, significantly enhancing the vulnerability detection capabilities of various intra-function features. This is particularly important in large and complex software systems where vulnerabilities often arise from intricate interactions among various components. The introduction of hypergraph-based modeling and using hypergraph neural networks to exploit multilateral associations provide a new frontier in vulnerability detection, offering a more comprehensive and effective solution for securing software systems against potential threats.

~\\
		\noindent
		\begin{minipage}[b]{0.45\textwidth}
			\fbox{
				\parbox{\textwidth}{
					\textbf{Answer to RQ2:} The experimental results indicate that compared to methods focusing solely on intra-function feature extraction, IFMA-VD significantly enhances vulnerability detection performance by constructing code behavior hypergraphs and utilizing hypergraph neural networks to extract multilateral association features.}
				}
			
		\end{minipage}
~\\

\subsection{\textbf{RQ3: Does IFMA-VD Effectively Detect the Real World Vulnerabilities?}}

 \begin{table*}[htbp]
  \centering
  \caption{Detection Accuracy of Different Methods on the Most Dangerous Vulnerabilities, \\Excluding Overlapping Data in Chrome+Debian (for RQ3).}

 \renewcommand{\arraystretch}{1.2}
   \setlength{\tabcolsep}{2pt}
  \begin{tabular}{c|c|c|c|c|c|c}
    \toprule
    \textbf{Rank} & \textbf{ID} & \textbf{Description} & \textbf{CodeBERT} & \textbf{VulBG} & \multicolumn{1}{c|}{\textbf{IFMA-VD}} & \textbf{Vuln} \\
    \midrule
    1     & CWE-787 & Out-of-bounds Write & 69    & 80    & \multicolumn{1}{c|}{\textbf{89}} & 136 \\
    2     & CWE-79 & Improper Neutralization of Input During Web Page Generation & 0     & \textbf{1} & \multicolumn{1}{c|}{0} & 1 \\
    3     & CWE-89 & Improper Neutralization of Special Elements used in an SQL Command & 2     & 1     & \multicolumn{1}{c|}{\textbf{1}} & 5 \\
    4     & CWE-416 & Use After Free & 92    & 109   & \multicolumn{1}{c|}{\textbf{128}} & 190 \\
    5     & CWE-78 & Improper Neutralization of Special Elements used in an OS Command & 3     & 5     & \multicolumn{1}{c|}{\textbf{6}} & 11 \\
    6     & CWE-20 & Improper Input Validation & 264   & 291   & \multicolumn{1}{c|}{\textbf{357}} & 658 \\
    7     & CWE-125 & Out-of-bounds Read & 183   & 231   & \multicolumn{1}{c|}{\textbf{375}} & 535 \\
    8     & CWE-22 & Improper Limitation of a Pathname to a Restricted Directory & 10    & \textbf{13} & \multicolumn{1}{c|}{10} & 22 \\
    9     & CWE-352 & Cross-Site Request Forgery & 0     & 0     & \multicolumn{1}{c|}{\textbf{1}} & 1 \\
    11    & CWE-862 & Missing Authorization & 0     & 1     & \multicolumn{1}{c|}{\textbf{1}} & 3 \\
    12    & CWE-476 & NULL Pointer Dereference & 82    & 113   & \multicolumn{1}{c|}{\textbf{129}} & 207 \\
    13    & CWE-287 & Improper Authentication & 10    & \textbf{12} & \multicolumn{1}{c|}{10} & 17 \\
    14    & CWE-190 & Integer Overflow or Wraparound & 89    & 101   & \multicolumn{1}{c|}{\textbf{128}} & 171 \\
    15    & CWE-502 & Deserialization of Untrusted Data & 0     & 0     & \multicolumn{1}{c|}{\textbf{1}} & 1 \\
    16    & CWE-77 & Improper Neutralization of Special Elements used in a Command & 3     & \textbf{4} & \multicolumn{1}{c|}{3} & 10 \\
    17    & CWE-119 & Improper Restriction of Operations within the Bounds of a Memory Buffer & 585   & 601   & \multicolumn{1}{c|}{\textbf{738}} & 1276 \\
    19    & CWE-918 & Server-Side Request Forgery & 0     & 1     & \multicolumn{1}{c|}{\textbf{1}} & 2 \\
    21    & CWE-362 & Concurrent Execution using Shared Resource with Improper Synchronization & 80    & 89    & \multicolumn{1}{c|}{\textbf{103}} & 182 \\
    22    & CWE-269 & Improper Privilege Management & 11    & \textbf{16} & \multicolumn{1}{c|}{13} & 29 \\
    23    & CWE-94 & Improper Control of Generation of Code  & 0     & \textbf{1} & \multicolumn{1}{c|}{0} & 3 \\
    24    & CWE-863 & Incorrect Authorization & 1     & 2     & \multicolumn{1}{c|}{\textbf{2}} & 4 \\
    \midrule
      -    &    -   & \textbf{Total} & 1484  & 1672  & \textbf{2096}  & 3464 \\
    \bottomrule
    \end{tabular}%
  \label{top}%
\end{table*}

To assess IFMA-VD's effectiveness in real world vulnerability detection, we measured its performance on the most critical and recent vulnerabilities. For comparison, we selected two baseline methods: the token-based CodeBERT and the code behavior graph-based VulBG.

$\bullet$ \textbf{  Effectiveness of IFMA-VD in Detecting the  Most Dangerous CWEs}

We measured the accuracy of IFMA-VD against the Common Weakness Enumeration (CWE) list of the most dangerous software vulnerabilities \cite{CWE25}. To construct a representative test dataset, we selected functions from the BigVul dataset \cite{fan2020ac} that correspond to these dangerous vulnerabilities, grouping them by CWE ID. As shown in Table \ref{top}, the BigVul dataset does not cover all the vulnerabilities in the list, resulting in a final dataset containing functions associated with 21 distinct dangerous CWE types. Table \ref{top} summarizes these types and the number of functions collected for each.

For model training, we used the Chrome+Debian dataset, selected due to its defect rate similarity to BigVul. To ensure a fair evaluation, we excluded any overlapping data from the test set. The IFMA-VD model with the highest F-measure from this training was then used to assess its detection accuracy.

The experimental results show that IFMA-VD detected 2,096 out of 3,463 vulnerabilities in the selected dangerous CWE types, achieving a detection rate of 60.5\%. This represents improvements of 41.2\% and 25.6\% over baseline methods. The results, as summarized in Table \ref{top}, demonstrate IFMA-VD's superior detection capability for most dangerous vulnerabilities in the evaluated CWE categories.

$\bullet$ \textbf{ Effectiveness of IFMA-VD in Detecting Newly Released Vulnerabilities}

To evaluate the effectiveness of IFMA-VD in detecting newly released vulnerabilities, we conducted experiments using recent vulnerability cases (Total 6) from 2023 to 2024 reported by FFmpeg Security \cite{ffmpeg}. These cases stem from the FFmpeg open-source project and cover commonly occurring vulnerability types in real-world scenarios, such as heap buffer overflow and out-of-bounds read. Detailed information is provided in the first four columns of Table \ref{ffmpeg}.

\begin{table*}[h]
  \centering

 \renewcommand{\arraystretch}{1.2}
  \caption{Detection Results of New Vulnerabilities in the FFmpeg Project 2023-2024 (for RQ3). }
    \begin{tabular}{c|c|c|c|c|c|c}
    \toprule
    \multirow{2}[4]{*}{\textbf{CVE ID}} & \multirow{2}[4]{*}{\textbf{CVulnerability Type}} & \multirow{2}[4]{*}{\textbf{Commit ID}} & \multirow{2}[4]{*}{\textbf{Filename}} & \multicolumn{3}{c}{\textbf{Detection Results}} \\
\cmidrule{5-7}    \multicolumn{1}{c|}{} & \multicolumn{1}{c|}{} & \multicolumn{1}{c|}{} & \multicolumn{1}{c|}{} & CodeBERT & VulBG & IFMA-VD \\
    \midrule
    CVE-2024-28661 & Heap Buffer Overflow & 66b50445 & speexdec.c & \multicolumn{1}{c|}{} &       & $\checkmark$ \\
    CVE-2024-7055 & Out-of-Bounds Access & d0ce2529 & pnmdec.c & $\checkmark$ & $\checkmark$ & $\checkmark$ \\
    CVE-2024-7272 & Out-of-Bounds Access & a937b3c5 & swresample.c & \multicolumn{1}{c|}{} &       & \multicolumn{1}{c}{} \\
    CVE-2023-47342 & Out-of-Bounds Access & e4d5ac8d & rtsp.c & \multicolumn{1}{c|}{} &       & $\checkmark$ \\
    CVE-2023-47343 & Illegal Memory Read & 0f6a3405 & mov.c & \multicolumn{1}{c|}{} & $\checkmark$ & $\checkmark$ \\
    CVE-2023-47344 & Out-of-Bounds Access & f7ac3512 & jpegxl\_parser.c & $\checkmark$ &       & $\checkmark$ \\
    \bottomrule
    \end{tabular}%
  \label{ffmpeg}%
\end{table*}

As shown in the last four columns of Table \ref{ffmpeg}, CodeBERT detected only two out-of-bounds access vulnerabilities, VulBG identified out-of-bounds access and illegal memory read. In contrast, the proposed IFMA-VD detected five vulnerabilities, including an additional heap buffer overflow, which the other two methods missed.

The comprehensive evaluation of IFMA-VD against the most dangerous vulnerabilities in the CWE  and the latest FFmpeg vulnerabilities demonstrates its superior capability in addressing real-world security threats. We attribute this performance improvement to IFMA-VD's integration of inter-function multilateral associations, which enhances the effectiveness of code feature extraction and, consequently, the model's vulnerability detection capabilities.

~\\
		\noindent
        \begin{minipage}[b]{0.45\textwidth}
			\fbox{
				\parbox{\textwidth}{
					\textbf{Answer to RQ3:} Experimental results indicate that the vulnerability detection model built with IFMA-VD outperforms the comparative methods in detecting real-world vulnerabilities. In conclusion, IFMA-VD is the preferred method for software reliability assurance teams to perform vulnerability detection tasks in most cases.}
				}
			
		\end{minipage}
~\\

\section{Discussion}

\subsection{How Effective Is IFMA-VD in Terms of Computational Resources?}

 Tables \ref{Train time} and \ref{Inference-time} present a comparative analysis of the training time, inference time, and model parameters for IFMA-VD and baseline methods. The results indicate that IFMA-VD achieves training times of only 12, 15, and 21 minutes, respectively, on the three datasets (excluding data processing time), significantly reducing training overhead. Compared to CodeBERT and PDBERT, this represents an improvement of approximately 60 times, and it outperforms other methods by several folds in terms of speed. Additionally, IFMA-DF demonstrates some of the best performance in terms of inference time and model parameter size. This efficiency is primarily attributed to the hypergraph's ability to compactly represent multilateral relationships among functions, thereby reducing the complexity of function behavior modeling, which in turn lowers training costs and model size. The short training time of IFMA-VD enables efficient fine-tuning and hyperparameter optimization across different projects, delivering superior performance, whereas the computational cost for other methods would be substantially higher. Furthermore, its smaller parameter size and reduced inference time make IFMA-DF particularly suitable for deployment on resource-constrained computing platforms.

 \begin{table}[htbp]
  \centering
   \setlength{\tabcolsep}{12pt}
   \caption{Train time of IFMA-VD v.s. Baseline Methods.}
   \renewcommand{\arraystretch}{1.1}
    \begin{tabular}{c|ccc}
    \toprule
    \textbf{Method} & \multicolumn{3}{c}{\textbf{Train time}} \\
    \midrule
    \multicolumn{1}{c|}{\textbf{Dataset}} & \textbf{FFmpeg} & \textbf{Qemu}  & \textbf{Chrome+Debian} \\
    \midrule
    Devign & 2h8m  & 2h39m & 3h11m \\
    CodeBERT & 20h15m & 26h37m & 34h5m \\
    IVDect & 2h35m & 3h7m  & 4h12m \\
    Reveal & 7h28m & 8h12m & 9h20m \\
    VulCNN & 50m   & 56m   & 1h18m \\
    VulBG & 56m   & 1h20m & 1h57m \\
    PDBERT & 21h23m & 27h15m & 34h52m \\
    \midrule
    IFMA-VD & 12m   & 15m   & 21m \\
    \bottomrule
    \end{tabular}%
  \label{Train time}%
\end{table}%

\begin{table}[htbp]
  \centering
     \setlength{\tabcolsep}{6pt}
  \caption{Inference time of IFMA-VD v.s. Baseline Methods.}
  \renewcommand{\arraystretch}{1.1}
    \begin{tabular}{c|ccc|c}
    \toprule
    \textbf{Method} & \multicolumn{3}{c|}{\textbf{Inference time}} & \# \textbf{parameters} \\
    \midrule
    \textbf{Dataset} & \textbf{FFmpeg} & \textbf{Qemu}  & \textbf{Chrome+Debia}n & - \\
    \midrule
    Devign & 4.1s  & 4.3s  & 5.5s  & 4.38MB \\
    CodeBERT & 5m    & 5m    & 6m    & 475MB \\
    IVDect & 3.9s  & 4.0s  & 4.6s  & 6.23MB \\
    Reveal & 2.3s  & 2.4s  & 3.1s  & 2.14MB \\
    VulCNN & 5.1s  & 5.4s  & 8.2s  & 2.6MB \\
    VulBG & 7.3s  & 8.7s  & 11s   & 24.1MB \\
    PDBERT & 5m    & 6m    & 6m    & 498MB \\
    IFMA-VD & 2.5 s  & 2.7s  & 3.0s  & 9.5MB \\
    \bottomrule
    \end{tabular}%
  \label{Inference-time}%
\end{table}%

\subsection{How Do Different Node Embedding Techniques and the Number of Hyperedges $K$ Affect the Experiment?}

IFMA-VD utilizes program slicing and node embedding techniques to extract code behaviors and filter out common behaviors to construct hyperedges. This section discusses the impact of varying embedding techniques and the number of hyperedges $K$ under the IFMA-VD method to achieve optimal results.

IFMA-VD utilizes embedding techniques to convert node tokens into vector representations during the generation of function CPGs and sliced subgraphs. As shown in Table \ref{embeding}, we experimented with three different embedding techniques: CodeBERT, FastText, and Word2Vec, across three datasets. The results indicate that Word2Vec provides the best performance, and thus, we chose Word2Vec as the node embedding technique.

\begin{table}[hpbt]
  \centering
  \setlength{\tabcolsep}{2pt}
  \caption{Performance of IFMA-VD Using Different Graph Node Embedding Techniques.}
  \renewcommand{\arraystretch}{1.1} 
    \begin{tabular}{c|cc|cc|cc}
    \toprule
    \multirow{2}[3]{*}{\textbf{Technique}} & \multicolumn{2}{c|}{\textbf{FFmpeg}} & \multicolumn{2}{c|}{\textbf{Qemu}} & \multicolumn{2}{c}{\textbf{Chrome+Debian}} \\
    \cmidrule{2-7}    & \textbf{F-measure}     & \textbf{Recall}     & \textbf{F-measure}     & \textbf{Recall}     & \textbf{F-measure}     & \textbf{Recall} \\
    \midrule
    CodeBERT &  66.5  & 85.1  & 58.6  & 67.2  & 39.7  & 59.8 \\

    FastText & 67.2  & 92.6  & 61.1  & 70.1  & 40.5  & 58.3 \\

    Word2Vec & \textbf{69.1} & \textbf{95.9} & \textbf{62.5} & \textbf{74.4} & \textbf{50.8} & \textbf{78.6} \\
    \bottomrule
    \end{tabular}%
  \label{embeding}%
\end{table}

\begin{figure}[h]
    \centerline{\includegraphics[width=0.9\linewidth]{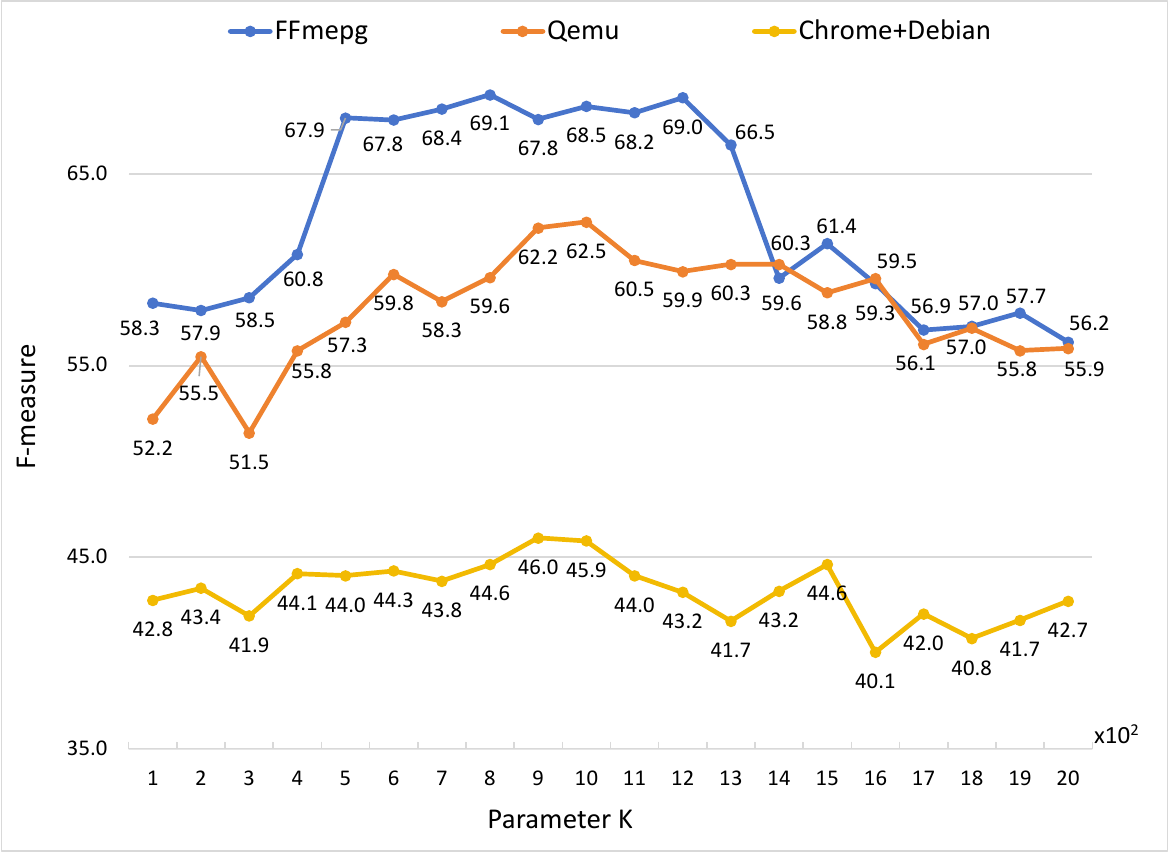}}
    \caption{The F-Measures of IFMA-VD Under Different Parameter K.}
    \label{K}
\end{figure}

For the number of hyperedges $K$, we only use $K$ as the variable in this experiment. Figure~\ref{K} shows the F-measure performance of IFMA-VD with different $K$ values on the three datasets. The performance peaks on the FFmpeg and Qemu datasets when $K$ is set to 800 and 1,000, respectively. Additionally, for Chrome+Debian, the F-measure performance is optimal when $K$ is set to 900. The results suggest that while the optimal $K$ value might vary slightly depending on the dataset, a value of around 1,000 generally yields the optimal results across diverse datasets.

			
\subsection{Threats to Validity}


\textbf{Data Preprocessing Time:}  In this paper, we construct inter-functiona multilateral associations by parsing the CPG and leveraging program slicing techniques. Due to the need to parse complex structures such as the ASTs and PDGs, generating the CPGs and slices is time-intensive. In practical detection and inference scenarios, the average time required to generate the CPG for a single sample is approximately 4 seconds.

\textbf{Implementation of Compared Methods:} We implemented several comparison methods in the experiments (e.g., Devign and Reveal) using the publicly released source code. For baselines without accessible source code, implementations were meticulously crafted based on the original papers. Nonetheless, our implementation may not capture all the nuances of the compared methods. 
		
\textbf{Experimental Results Might Not Be Generalizable:}
To assess the generalizability of IFMA-VD, we conducted experiments on diverse open-source vulnerability detection datasets exhibiting varying vulnerability numbers and rates. While these findings demonstrate IFMA-VD's effectiveness, existing research suggests that some datasets might contain noise \cite{croft2023data}. Therefore, further research is necessary to evaluate its performance across a broader spectrum of vulnerability datasets to mitigate threats to external validity.

\textbf{F-measure and Recall Might Not Be the Only Appropriate Metrics:} We utilized F-measure and Recall as our evaluation metrics because these are commonly used in vulnerability detection research. Different metrics may yield different results. This paper does not include comparisons with other metrics (e.g., PR-AUC and Accuracy). Future work should consider incorporating a broader range of evaluation metrics.

\section{Conclusion}
In this paper, we propose IFMA-VD to unveil multilateral association characteristics in vulnerability detection. IFMA-VD enhances the effectiveness of code intra-function features by transforming function encoding into a CPG. Additionally, it employs program slicing to construct hypergraphs of code behaviour, aiming to mine multilateral associations and thereby improve the effectiveness of the modelling data. Empirical studies indicate that the vulnerability detection models built using features extracted by IFMA-VD achieve superior F-measure and Recall results on experiment datasets. In most scenarios, we believe that IFMA-VD could be a viable option for vulnerability detection tasks. 


Several issues remain for future work. Firstly, we plan to incorporate more software code data, including open-source and commercial projects, to validate our method's performance further. Secondly, we aim to enhance the effectiveness of IFMA-VD by integrating additional code modals, specifically modals based on structure and visualization.

\bibliographystyle{ieeetr}
\normalem
\bibliography{main}

\begin{thebibliography}{10}

\bibitem{lin2020software}
G.~Lin, S.~Wen, Q.-L. Han, J.~Zhang, and Y.~Xiang, ``Software vulnerability
  detection using deep neural networks: a survey,'' {\em Proceedings of the
  IEEE}, vol.~108, no.~10, pp.~1825--1848, 2020.

\bibitem{jiang2024stagedvulbert}
Y.~Jiang, Y.~Zhang, X.~Su, C.~Treude, and T.~Wang, ``Stagedvulbert:
  Multi-granular vulnerability detection with a novel pre-trained code model,''
  {\em IEEE Transactions on Software Engineering}, 2024.

\bibitem{NVD2023}
{National vulnerability database}. \url{https://nvd.nist.gov/}, 2023.

\bibitem{ni2023function}
C.~Ni, X.~Guo, Y.~Zhu, X.~Xu, and X.~Yang, ``Function-level vulnerability
  detection through fusing multi-modal knowledge,'' in {\em 2023 38th IEEE/ACM
  International Conference on Automated Software Engineering (ASE)},
  pp.~1911--1918, IEEE, 2023.

\bibitem{nong2022open}
Y.~Nong, R.~Sharma, A.~Hamou-Lhadj, X.~Luo, and H.~Cai, ``Open science in
  software engineering: A study on deep learning-based vulnerability
  detection,'' {\em IEEE Transactions on Software Engineering}, vol.~49, no.~4,
  pp.~1983--2005, 2022.

\bibitem{li2021sysevr}
Z.~Li, D.~Zou, S.~Xu, H.~Jin, Y.~Zhu, and Z.~Chen, ``Sysevr: A framework for
  using deep learning to detect software vulnerabilities,'' {\em IEEE
  Transactions on Dependable and Secure Computing}, vol.~19, no.~4,
  pp.~2244--2258, 2021.

\bibitem{guo2022unixcoder}
D.~Guo, S.~Lu, N.~Duan, Y.~Wang, M.~Zhou, and J.~Yin, ``Unixcoder: Unified
  cross-modal pre-training for code representation,'' in {\em Proceedings of
  the 60th Annual Meeting of the Association for Computational Linguistics
  (Volume 1: Long Papers)}, pp.~7212--7225, 2022.

\bibitem{zhang2023vulnerability}
J.~Zhang, Z.~Liu, X.~Hu, X.~Xia, and S.~Li, ``Vulnerability detection by
  learning from syntax-based execution paths of code,'' {\em IEEE Transactions
  on Software Engineering}, 2023.

\bibitem{wen2023meta}
X.-C. Wen, C.~Gao, J.~Ye, Y.~Li, Z.~Tian, Y.~Jia, and X.~Wang, ``Meta-path
  based attentional graph learning model for vulnerability detection,'' {\em
  IEEE Transactions on Software Engineering}, 2023.

\bibitem{zhang2024evm}
X.~Zhang, W.~Sun, Z.~Xu, H.~Cheng, C.~Cai, H.~Cui, and Q.~Li, ``Evm-shield:
  In-contract state access control for fast vulnerability detection and
  prevention,'' {\em IEEE Transactions on Information Forensics and Security},
  2024.

\bibitem{yuan2023enhancing}
B.~Yuan, Y.~Lu, Y.~Fang, Y.~Wu, D.~Zou, Z.~Li, Z.~Li, and H.~Jin, ``Enhancing
  deep learning-based vulnerability detection by building behavior graph
  model,'' in {\em 2023 IEEE/ACM 45th International Conference on Software
  Engineering (ICSE)}, pp.~2262--2274, IEEE, 2023.

\bibitem{yadavally2024learning}
A.~Yadavally, Y.~Li, S.~Wang, and T.~N. Nguyen, ``A learning-based approach to
  static program slicing,'' {\em Proceedings of the ACM on Programming
  Languages}, vol.~8, no.~OOPSLA1, pp.~83--109, 2024.

\bibitem{xia2021graph}
F.~Xia, K.~Sun, S.~Yu, A.~Aziz, L.~Wan, S.~Pan, and H.~Liu, ``Graph learning: A
  survey,'' {\em IEEE Transactions on Artificial Intelligence}, vol.~2, no.~2,
  pp.~109--127, 2021.

\bibitem{zhang2022rethinking}
B.~Zhang, S.~Luo, L.~Wang, and D.~He, ``Rethinking the expressive power of gnns
  via graph biconnectivity,'' in {\em The Eleventh International Conference on
  Learning Representations}, 2022.

\bibitem{yang2021graphformers}
J.~Yang, Z.~Liu, S.~Xiao, C.~Li, D.~Lian, S.~Agrawal, A.~Singh, G.~Sun, and
  X.~Xie, ``Graphformers: Gnn-nested transformers for representation learning
  on textual graph,'' {\em Advances in Neural Information Processing Systems},
  vol.~34, pp.~28798--28810, 2021.

\bibitem{di2022generating}
D.~Di, C.~Zou, Y.~Feng, H.~Zhou, R.~Ji, Q.~Dai, and Y.~Gao, ``Generating
  hypergraph-based high-order representations of whole-slide histopathological
  images for survival prediction,'' {\em IEEE Transactions on Pattern Analysis
  and Machine Intelligence}, vol.~45, no.~5, pp.~5800--5815, 2022.

\bibitem{feng2023hypergraph}
Y.~Feng, S.~Ji, Y.-S. Liu, S.~Du, Q.~Dai, and Y.~Gao, ``Hypergraph-based
  multi-modal representation for open-set 3d object retrieval,'' {\em IEEE
  Transactions on Pattern Analysis and Machine Intelligence}, 2023.

\bibitem{cheng2021deepwukong}
X.~Cheng, H.~Wang, J.~Hua, G.~Xu, and Y.~Sui, ``Deepwukong: Statically
  detecting software vulnerabilities using deep graph neural network,'' {\em
  ACM Transactions on Software Engineering and Methodology (TOSEM)}, vol.~30,
  no.~3, pp.~1--33, 2021.

\bibitem{mamede2022transformer}
C.~Mamede, E.~Pinconschi, and R.~Abreu, ``A transformer-based ide plugin for
  vulnerability detection,'' in {\em Proceedings of the 37th IEEE/ACM
  International Conference on Automated Software Engineering}, pp.~1--4, 2022.

\bibitem{lipp2022empirical}
S.~Lipp, S.~Banescu, and A.~Pretschner, ``An empirical study on the
  effectiveness of static c code analyzers for vulnerability detection,'' in
  {\em Proceedings of the 31st ACM SIGSOFT International Symposium on Software
  Testing and Analysis}, pp.~544--555, 2022.

\bibitem{wu2024ultravcs}
T.~Wu, L.~Chen, G.~Du, D.~Meng, and G.~Shi, ``Ultravcs: Ultra-fine-grained
  variable-based code slicing for automated vulnerability detection,'' {\em
  IEEE Transactions on Information Forensics and Security}, 2024.

\bibitem{afrose2022evaluation}
S.~Afrose, Y.~Xiao, S.~Rahaman, B.~P. Miller, and D.~Yao, ``Evaluation of
  static vulnerability detection tools with java cryptographic api
  benchmarks,'' {\em IEEE Transactions on Software Engineering}, vol.~49,
  no.~2, pp.~485--497, 2022.

\bibitem{xue2020cross}
Y.~Xue, M.~Ma, Y.~Lin, Y.~Sui, J.~Ye, and T.~Peng, ``Cross-contract static
  analysis for detecting practical reentrancy vulnerabilities in smart
  contracts,'' in {\em Proceedings of the 35th IEEE/ACM International
  Conference on Automated Software Engineering}, pp.~1029--1040, 2020.

\bibitem{liu2024semantic}
Y.~Liu, C.~Zhang, F.~Li, Y.~Li, J.~Zhou, J.~Wang, L.~Zhan, Y.~Liu, and W.~Huo,
  ``Semantic-enhanced static vulnerability detection in baseband firmware,'' in
  {\em Proceedings of the IEEE/ACM 46th International Conference on Software
  Engineering}, pp.~1--12, 2024.

\bibitem{zhu2022fuzzing}
X.~Zhu, S.~Wen, S.~Camtepe, and Y.~Xiang, ``Fuzzing: a survey for roadmap,''
  {\em ACM Computing Surveys (CSUR)}, vol.~54, no.~11s, pp.~1--36, 2022.

\bibitem{zheng2022park}
P.~Zheng, Z.~Zheng, and X.~Luo, ``Park: Accelerating smart contract
  vulnerability detection via parallel-fork symbolic execution,'' in {\em
  Proceedings of the 31st ACM SIGSOFT International Symposium on Software
  Testing and Analysis}, pp.~740--751, 2022.

\bibitem{zhang2022example}
Y.~Zhang, Y.~Xiao, M.~M.~A. Kabir, D.~Yao, and N.~Meng, ``Example-based
  vulnerability detection and repair in java code,'' in {\em Proceedings of the
  30th IEEE/ACM International Conference on Program Comprehension},
  pp.~190--201, 2022.

\bibitem{chen2022ausera}
S.~Chen, Y.~Zhang, L.~Fan, J.~Li, and Y.~Liu, ``Ausera: Automated security
  vulnerability detection for android apps,'' in {\em Proceedings of the 37th
  IEEE/ACM International Conference on Automated Software Engineering},
  pp.~1--5, 2022.

\bibitem{zou2022mvulpreter}
D.~Zou, Y.~Hu, W.~Li, Y.~Wu, H.~Zhao, and H.~Jin, ``mvulpreter: A
  multi-granularity vulnerability detection system with interpretations,'' {\em
  IEEE Transactions on Dependable and Secure Computing}, 2022.

\bibitem{qiu2024vulnerability}
F.~Qiu, Z.~Liu, X.~Hu, X.~Xia, G.~Chen, and X.~Wang, ``Vulnerability detection
  via multiple-graph-based code representation,'' {\em IEEE Transactions on
  Software Engineering}, 2024.

\bibitem{huang2024hevuld}
Y.~Huang, M.~He, X.~Wang, and J.~Zhang, ``Hevuld: A static vulnerability
  detection method using heterogeneous graph code representation,'' {\em IEEE
  Transactions on Information Forensics and Security}, 2024.

\bibitem{cao2024coca}
S.~Cao, X.~Sun, X.~Wu, D.~Lo, L.~Bo, B.~Li, and W.~Liu, ``Coca: Improving and
  explaining graph neural network-based vulnerability detection systems,'' in
  {\em Proceedings of the IEEE/ACM 46th International Conference on Software
  Engineering}, pp.~1--13, 2024.

\bibitem{cao2022mvd}
S.~Cao, X.~Sun, L.~Bo, R.~Wu, B.~Li, and C.~Tao, ``Mvd: memory-related
  vulnerability detection based on flow-sensitive graph neural networks,'' in
  {\em Proceedings of the 44th international conference on software
  engineering}, pp.~1456--1468, 2022.

\bibitem{li2021vulnerability}
Y.~Li, S.~Wang, and T.~N. Nguyen, ``Vulnerability detection with fine-grained
  interpretations,'' in {\em Proceedings of the 29th ACM Joint Meeting on
  European Software Engineering Conference and Symposium on the Foundations of
  Software Engineering}, pp.~292--303, 2021.

\bibitem{hu2023interpreters}
Y.~Hu, S.~Wang, W.~Li, J.~Peng, Y.~Wu, D.~Zou, and H.~Jin, ``Interpreters for
  gnn-based vulnerability detection: Are we there yet?,'' in {\em Proceedings
  of the 32nd ACM SIGSOFT International Symposium on Software Testing and
  Analysis}, pp.~1407--1419, 2023.

\bibitem{steenhoek2024dataflow}
B.~Steenhoek, H.~Gao, and W.~Le, ``Dataflow analysis-inspired deep learning for
  efficient vulnerability detection,'' in {\em Proceedings of the 46th IEEE/ACM
  International Conference on Software Engineering}, pp.~1--13, 2024.

\bibitem{liu2024pre}
Z.~Liu, Z.~Tang, J.~Zhang, X.~Xia, and X.~Yang, ``Pre-training by predicting
  program dependencies for vulnerability analysis tasks,'' in {\em Proceedings
  of the IEEE/ACM 46th International Conference on Software Engineering},
  pp.~1--13, 2024.

\bibitem{wu2022vulcnn}
Y.~Wu, D.~Zou, S.~Dou, W.~Yang, D.~Xu, and H.~Jin, ``Vulcnn: An image-inspired
  scalable vulnerability detection system,'' in {\em Proceedings of the 44th
  International Conference on Software Engineering}, pp.~2365--2376, 2022.

\bibitem{wen2023vulnerability}
X.-C. Wen, Y.~Chen, C.~Gao, H.~Zhang, J.~M. Zhang, and Q.~Liao, ``Vulnerability
  detection with graph simplification and enhanced graph representation
  learning,'' in {\em 2023 IEEE/ACM 45th International Conference on Software
  Engineering (ICSE)}, pp.~2275--2286, IEEE, 2023.

\bibitem{Devign}
Y.~Zhou, S.~Liu, J.~Siow, X.~Du, and Y.~Liu, ``Devign: Effective vulnerability
  identification by learning comprehensive program semantics via graph neural
  networks,'' {\em Advances in neural information processing systems}, vol.~32,
  2019.

\bibitem{qiu2024code}
S.~Qiu, M.~Huang, Y.~Liang, C.~Peng, and Y.~Yuan, ``Code multiview hypergraph
  representation learning for software defect prediction,'' {\em IEEE
  Transactions on Reliability}, 2024.

\bibitem{chakraborty2021deep}
S.~Chakraborty, R.~Krishna, Y.~Ding, and B.~Ray, ``Deep learning based
  vulnerability detection: Are we there yet?,'' {\em IEEE Transactions on
  Software Engineering}, vol.~48, no.~9, pp.~3280--3296, 2021.

\bibitem{feng2020CodeBERT}
Z.~Feng, D.~Guo, D.~Tang, N.~Duan, X.~Feng, M.~Gong, L.~Shou, B.~Qin, T.~Liu,
  D.~Jiang, {\em et~al.}, ``Codebert: A pre-trained model for programming and
  natural languages,'' {\em arXiv preprint arXiv:2002.08155}, 2020.

\bibitem{CWE25}
{CWE Top 25 Most Dangerous Software Weaknesses}.
  \url{https://cwe.mitre.org/top25/archive/2023/2023_top25_list.html}, 2023.

\bibitem{fan2020ac}
J.~Fan, Y.~Li, S.~Wang, and T.~N. Nguyen, ``Ac/c++ code vulnerability dataset
  with code changes and cve summaries,'' in {\em Proceedings of the 17th
  International Conference on Mining Software Repositories}, pp.~508--512,
  2020.

\bibitem{ffmpeg}
{The latest ffmpeg vulnerabilities released (2023-2024)}. \url{FFmpeg Security.
  https://www.ffmpeg.org/security.html}, 2024.

\bibitem{croft2023data}
R.~Croft, M.~A. Babar, and M.~M. Kholoosi, ``Data quality for software
  vulnerability datasets,'' in {\em 2023 IEEE/ACM 45th International Conference
  on Software Engineering (ICSE)}, pp.~121--133, IEEE, 2023.

\end{thebibliography}
\end{document}